\documentclass[aps, twocolumn, a4paper, superscriptaddress, 10pt, pre]{revtex4-2}

\usepackage{graphicx}
\usepackage{amsmath, amssymb}
\usepackage{lmodern}
\usepackage[T1]{fontenc}
\usepackage[utf8]{inputenc}
\usepackage[english]{babel}
\usepackage[dvipsnames]{xcolor}
\usepackage{lipsum}
\usepackage{booktabs}
\usepackage{changes}
\usepackage{pifont}
\usepackage[colorlinks=true, allcolors=blue]{hyperref}

\begin{document}

\title{Investigating Mobility in Spatial Biodiversity Models through Recurrence Quantification Analysis}

\author{M. S. Palmero}
\email{palmero@usp.com}
\affiliation{Instituto de Ciências Matemáticas e de Computação, Universidade de São Paulo, 13560-970 São Carlos, SP, Brazil}
\affiliation{Potsdam Institute for Climate Impact Research (PIK), Member of the Leibniz Association, Telegrafenberg A31, 14473 Potsdam, Germany}

\author{M. Bongestab}
\affiliation{Departamento de Física, Universidade Federal da Paraíba, 58051-970 João Pessoa, PB, Brazil}

\author{N. Marwan}
\affiliation{Potsdam Institute for Climate Impact Research (PIK), Member of the Leibniz Association, Telegrafenberg A31, 14473 Potsdam, Germany}

\newcommand{\psketch}{\scriptsize \color{RoyalPurple}}
\newcommand{\bsketch}{\scriptsize \color{violet}}

\newcommand{\amended}[1]{{\color{BrickRed}{#1}}}

\begin{abstract}
Recurrence plots and their associated quantifiers provide a robust framework for detecting and characterising complex patterns in non-linear time-series. In this paper, we employ recurrence quantification analysis to investigate the dynamics of the cyclic, non-hierarchical May-Leonard model, also referred to as rock--paper--scissors systems, that describes competitive interactions among three species. A crucial control parameter in these systems is the species' mobility $m$, which governs the spatial displacement of individuals and profoundly influences the resulting dynamics. By systematically varying $m$ and constructing suitable recurrence plots from numerical simulations, we explore how recurrence quantifiers reflect distinct dynamical features associated with different ecological states. We then introduce an ensemble-based approach that leverages statistical distributions of recurrence quantifiers, computed from numerous independent realisations, allowing us to identify dynamical outliers as significant deviations from typical system behaviour. Through detailed numerical analyses, we demonstrate that these outliers correspond to divergent ecological regimes associated with specific mobility values, providing also a robust manner to infer the mobility parameter from observed numerical data. Our results highlight the potential of recurrence-based methods as diagnostic tools for analysing spatial ecological systems and extracting ecologically relevant information from their non-linear dynamical patterns.
\end{abstract}

\maketitle

\section{Introduction}
\label{intro}

The question of how biodiversity is maintained in ecological systems has long captivated researchers in theoretical ecology \cite{leveque2003biodiversity, nowak2006evolutionary}. In spatially structured populations, non-hierarchical interactions between species, as captured by the May--Leonard model and often studied through the rock--paper--scissors (RPS) framework, have become a widely used paradigm for understanding mechanisms of coexistence \cite{Frean2001, Kerr2002}. Its simplicity has been confirmed to stabilize diversity in several other contexts \cite{Jackson1975,sinervo1996rock, kirkup2004antibiotic, ruifrok2015cyclical, grace2015regulation, perc2017statistical, shibasaki2018cyclic, xu2023interacting}, in particular, in the recent study using strains of \textit{E. Coli} in an in--vitro experiment Ref. \cite{liao2020survival}. In this game, each species dominates another in a uni--directed closed cyclic loop, which prevents the emergence of a strict competitive hierarchy. When spatial effects are taken into account and interactions are restricted to local neighbourhoods, RPS dynamics can give rise to spatio-temporal structures, such as spiral and front--waves, that are linked to long--term coexistence across the system \cite{szolnoki2014cyclic, szolnoki2020pattern, bazeia2022influence}.

One of the key parameters governing the system's behaviour is the species' mobility, denoted by $m$, which determines the rate at which individuals move across the spatial domain. Changes in $m$ can induce different dynamical regimes. At low mobility, the system self-organises into spatial domains that maintain biodiversity. In contrast, at higher mobility values, these spatial structures gradually disappear, extinction events become more frequent, spatial correlations weaken, and biodiversity ultimately collapses as a single species dominates the system \cite{reichenbach2007mobility, bazeia2017hamming}. Understanding the transition of these antagonistic dynamical regimes as a function of the mobility parameter, especially in the presence of stochastic fluctuations, remains an important challenge in theoretical ecology.

To address this challenge, we adopt a methodological framework grounded in non-linear time-series analysis. Recurrence Plots (RPs) were originally introduced as graphical tools for visualising the recurrence of states in dynamical systems \cite{Eckmann1987}, and have since turned into a robust and versatile method for characterising complex systems \cite{marwan2007recurrence, Webber2005, marwan2008epjst}. Their quantitative extension, known as Recurrence quantification analysis (RQA), involves deriving numerical indicators directly from the recurrence structures, capturing essential characteristics of the underlying dynamics \cite{marwan2007recurrence,MarwanWebber2015}. These recurrence-based measures have demonstrated effectiveness in distinguishing periodic, quasi-periodic, chaotic, and stochastic signals across various scientific fields. Applications of recurrence analysis have become widespread, spanning areas such as physics \cite{Thiel2002,Zolotova2006,Palmero2023}, physiology and neuroscience \cite{Marwan2002,Prabhu2020, Acharya2011}, climate and earth sciences \cite{Donges2011,spiridonov2017,Trauth2019}, finance and economics \cite{Bastos2011, Fabretti2005}, engineering systems \cite{Nichols2006, Sen2008,kasthuri2019}, and even in the context of ecological research \cite{semeraro2021,semeraro2020,zurlini2018,ayers2015}.

In this work, we investigate how recurrence analysis may enhance our understanding of the role played by the mobility parameter in spatial biodiversity models, specifically focusing on the spatial May–Leonard or RPS system. Mobility is known to influence the formation of spatio-temporal patterns and is crucial both for maintaining and threatening biodiversity. As shown by \citet{reichenbach2007mobility}, biodiversity collapses beyond a critical mobility threshold. Therefore, our study concentrates on the complex temporal fluctuations of species' abundances within the coexistence regime. We apply RQA to species' abundance time-series derived from numerical simulations of the RPS model under varying mobility conditions. Rather than relying on individual trajectory analysis, we adopt an ensemble-based approach, similar to those proposed by \cite{Palmero2023, braun2023}, to construct statistical distributions of recurrence quantifiers for each mobility value from multiple independent realizations.

The proposed methodology aims to explore a full range of dynamical behaviours present in the system’s coexistence phase, including atypical or outlier dynamics. Identifying such atypical cases allows us to detect subtle but significant shifts in ecological processes, which may precede regime transitions associated with variations in mobility. The approach provides a rigorous framework to distinguish between common and anomalous ecological fluctuations, and it introduces a novel strategy for inferring ecological parameters, such as mobility $m$, directly from empirical time-series of species abundance. This methodology is particularly well suited to ecological systems that exhibit cyclic competitive interactions. Examples may include microbial communities structured by toxin-mediated inhibition \cite{Kerr2002}, territorial lizard populations with cyclical dominance of colour morphs \cite{sinervo1996rock}, and marine invertebrate assemblages governed by localised competition \cite{Jackson1975}. Ultimately, these methodological developments can support more nuanced analyses and promote improved interpretation of observational ecological data.

The remainder of the paper is organised as follows. Section~\ref{sec:model} presents the details of the spatial RPS model and outlines the proposed recurrence-based methodology. In Sec.~\ref{sec:results}, we discuss the results obtained from extensive numerical simulations, examining how recurrence quantifiers vary with the mobility parameter. Additionally, we describe our approach for detecting and inferring outlier mobility regimes, while also providing a comparison to conventional time-series analysis in Appendix~\ref{appendix_A}. Finally, Sec.~\ref{sec:conclude} summarises the main contributions of this work and outlines potential directions for future research.

\section{Model and Methodology}
\label{sec:model}

This section aims to outline and explain both the model and the methodology used in this work. The first subsection provides a detailed description of the May-Leonard model for the cyclic competition among three distinct species. The second subsection focuses on the recurrence-based methodology employed to analyse the complex dynamics emerging from the May-Leonard model, presenting the general construction of a recurrence plot and its quantification analysis.

\subsection{Rock--Paper--Scissor Model}
\label{subsec:model}

Here we introduce the afore-mentioned population dynamics model with three cyclically competing species \cite{szolnoki2020pattern, bazeia2021effects, bazeia2022influence, bazeia2024chaotic}. 

\begin{figure}[htpb!]
    \centering
    \includegraphics[scale=1]{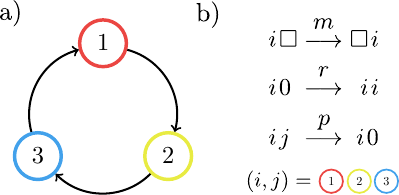}
    \caption{a) Cyclical non-hierarchical predation chain of the RPS game. Individuals from species $1$ outcompetes individuals from species $2$, while the ones from $2$ outcompetes the individuals from $3$, lastly individuals from species $3$ outcompetes individuals of species $1$, closing the loop; b) Reaction scheme of the local rules of mobility $(m)$, reproduction $(r)$ and predation $(p)$}
    \label{fig:rps+rules}
\end{figure}

The model under investigation is the well--known three--species May--Leonard (ML) model, also known as rock--paper--scissors, RPS. The cyclic non-hierarchical competition feature among the species leads to a dynamical coexistence of the participating species \cite{may1975nonlinear} and the occurrence of spiral patterns in the spatial version of the time evolution \cite{reichenbach2007mobility, szolnoki2020pattern}. At the individual level, the participants of each species are allowed to interact, with a given probability, within its first Von--Neumann neighbours (up, down, left, and right) in three distinct ways, via: mobility (m), reproduction (r) and predation (p). The number of free variables can be reduced, without loss of generality, by setting $r=p=(1-m)/2$ while conserving $m+r+p=1$. In particular, in the competition action the cyclical non-hierarchical dominance is implemented as illustrated by Fig.~\ref{fig:rps+rules} a).

\begin{figure*}[t!]
    \centering
    \includegraphics[scale=1]{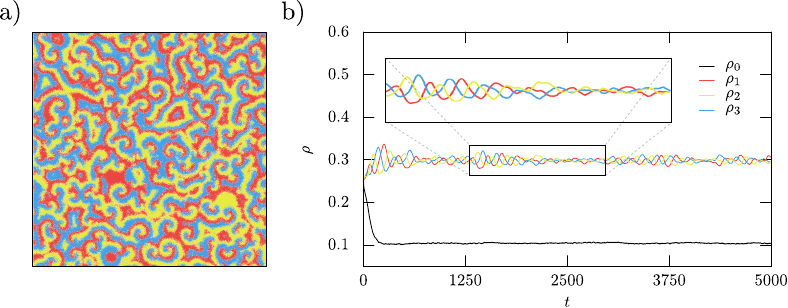}
    \caption{a) Final state of a single simulations in a lattice sized $N = 10^3$ for $m=0.60$ and $r=p=0.20$. The red colour represent the species 1 individuals, the yellow species 2 and the blue colour the individuals from species 3; b) Characteristic abundance time-series for each occupation in the lattice, evolved up to 5,000 generations. The system reaches the relaxed state after a first transient moment leaving the well mixed state to the dynamical spiral patterns. At the relaxed state, the empty spaces occupy around $10\%$ of the total lattice area. The remaining part is almost equally occupied by the competing species, around $30\%$. The inset shows the complex oscillatory dynamics of the competing species.}
    \label{fig:dst}
\end{figure*}

The population dynamics happens in a regular square lattice of size $N \times N$. For the initial condition we prepare a homogeneous state by randomly spreading individuals of species 1, 2, and 3 and also empty spaces represented by 0 throughout the $N^2$ sites, this way no preferential advantage is given for any species. The system evolves in time following the sequential steps: first a random site $i$ is chosen, it is called the active one; in the same manner, a site $j$ among its neighbours is also randomly selected, it is the passive one. An empty site can never be an active site. In the sequence, it is decided how the active site will act upon the passive one, therefore one of the three actions is randomly selected with its associated probability $(m,r,p)$: {\bf (1)} If the mobility action is chosen the positions of the active and passive sites are swapped $(i,\square) \rightarrow (\square,i)$, independent of the occupation in the passive site. If the mobility action was not chosen, then either of the next two actions are considered: {\bf (2)} If the reproduction rule is selected and the passive site is empty, it will be filled with one individual of the active site $(i,0) \rightarrow (i,i)$. {\bf (3)} Finally, if competition is to be implemented the active and passive individual must satisfy the RPS rule and we have $(i,j) \rightarrow (i,0)$, creating and empty site. Figure~\ref{fig:rps+rules} b) illustrates these local rules for the $(m,r,p)$ parameters.

It is also important to stress that this is an individual based model, where only one action at a time is attempted for execution, and it is always executed by the active site if the conditions for implementation are satisfied. This is referred as a successful interaction. After $N^2$ successful interactions between a random active site $i$ with a passive neighbour $j$ throughout the lattice, a generation is recorded and the data from the state of the simulation can be collected. 

Two important output information from the collected data are the final state of the system as depicted in Fig.~\ref{fig:dst} a), and the time-series of species' abundance, represented here by $\rho_i(t)$ with $i=0,1,2,$ and $3$ presented in Fig.~\ref{fig:dst} b), evolved up to $t=T=5,000$, where $T$ is a predefined maximum iteration time. In both panels, the colour red is associated with species 1, colour yellow with species 2 and the colour blue with species 3. Moreover, in a typical time evolution of the ML model, the system starts from the homogeneous initial state with $\rho_i=1/4$, and after a transient time, the species abundances oscillates around a well defined average value and so does the empty sites, but with a smaller amplitude.



For the particular case in Fig.~\ref{fig:dst}, the parameters of the simulations are $m=0.5$, $r=p=0.25$. In the spatial ML system, mobility plays a critical role both in sustaining the characteristic spiral wave dynamics and in regulating biodiversity. It is known that there exists a critical mobility threshold, which depends on the lattice size, beyond which biodiversity is lost \cite{reichenbach2007mobility}. When mobility exceeds this threshold, the system inevitably evolves toward the dominance of a single species, thereby collapsing the dynamical coexistence. In this study, our focus is to investigate the coexistence phase in detail. To that end, we adopt a large lattice size ($N = 10^3$) in order to minimise the occurrence of extinction events. For this choice, the critical mobility is $m_c = 0.998$, and our analysis is restricted to values of $m$ within the interval $[0.000, 0.998]$.

\subsection{Recurrence Plot Analysis}
\label{subsec:methodology}

The representation of the recurrence structure of a dynamical system by a recurrence plot (RP) is a powerful technique for uncovering hidden structures in complex signals. Since many real-world problems provide only a single observed variable, an effective way to analyse their underlying dynamics is through time-delay embedding, a method inspired by Takens' embedding theorem \cite{Takens1981, Noakes1991}. This approach reconstructs a 
$d$-dimensional representation of the signal by introducing delayed copies of itself. Given a scalar time-series $x(t)$, the embedded representation is constructed as

\begin{equation}
    \mathbf{X}(t) = \left[ x(t), x(t + \tau), x(t + 2\tau), \dots, x(t + (d-1)\tau) \right],
\end{equation}where $\tau$ is the time delay and $d$ is the embedding dimension. Setting $\tau = 1$ and $d = 2$ results in a two-dimensional representation where each state vector consists of two consecutive points from the time-series, $\mathbf{X}(t) = [x(t), x(t+1)]$. This simple embedding captures the immediate relationship between consecutive states, which is particularly suitable for the iterative dynamics of the RPS system analysed here, where the effective numerical iteration is one generation. Nonetheless, alternative embedding parameters derived from more robust methods \cite{kraemer2021, Marwan2023} may yield complementary insights, depending on the specific features or questions one aims to explore.

Once the time-series is embedded, a RP is constructed by identifying recurrent states through the recurrence matrix (RM), a binary square matrix composed by the elements $R_{i,j}$, $\mathbf{R} \in \{0, 1\}^{T \times T}$, where $T$ is the total iteration time, as follows
\begin{equation}
    R_{ij} = \Theta(\varepsilon - \|\mathbf{X}_i - \mathbf{X}_j\|),
\end{equation}where $\varepsilon$ is a predefined distance threshold, $\Theta(\cdot)$ is the Heaviside step function, and $\|\cdot\|$ is a suitable norm. The Euclidean norm is used throughout this study. The RM reveals temporal structures and recurring patterns within the signal, enabling further analysis of its complexity and dynamical properties.

To allow consistent comparison across different simulations, all RPs in this work are constructed with a fixed recurrence rate (RR), rather than a fixed threshold \cite{kraemer2018}. The process begins by computing the pairwise distance matrix $\mathbf{D} \in \mathbb{R}^{T \times T}$, where each element is given by
\begin{equation}
    D_{ij} = \| \mathbf{X}_i - \mathbf{X}_j \|.
    \label{eq:distance_matrix}
\end{equation}To impose the desired RR, the threshold $\varepsilon$ is selected such that the proportion of points satisfying $D_{ij} \leq \varepsilon$ equals RR (RR-percentile):
\begin{equation}
    \frac{1}{N^2} \sum_{i=1}^N \sum_{j=1}^N \Theta(\varepsilon - D_{ij}) = \mathrm{RR}.
    \label{eq:requirement_percentage}
\end{equation}

\begin{figure}[h!]
    \centering
    \includegraphics[scale=0.43]{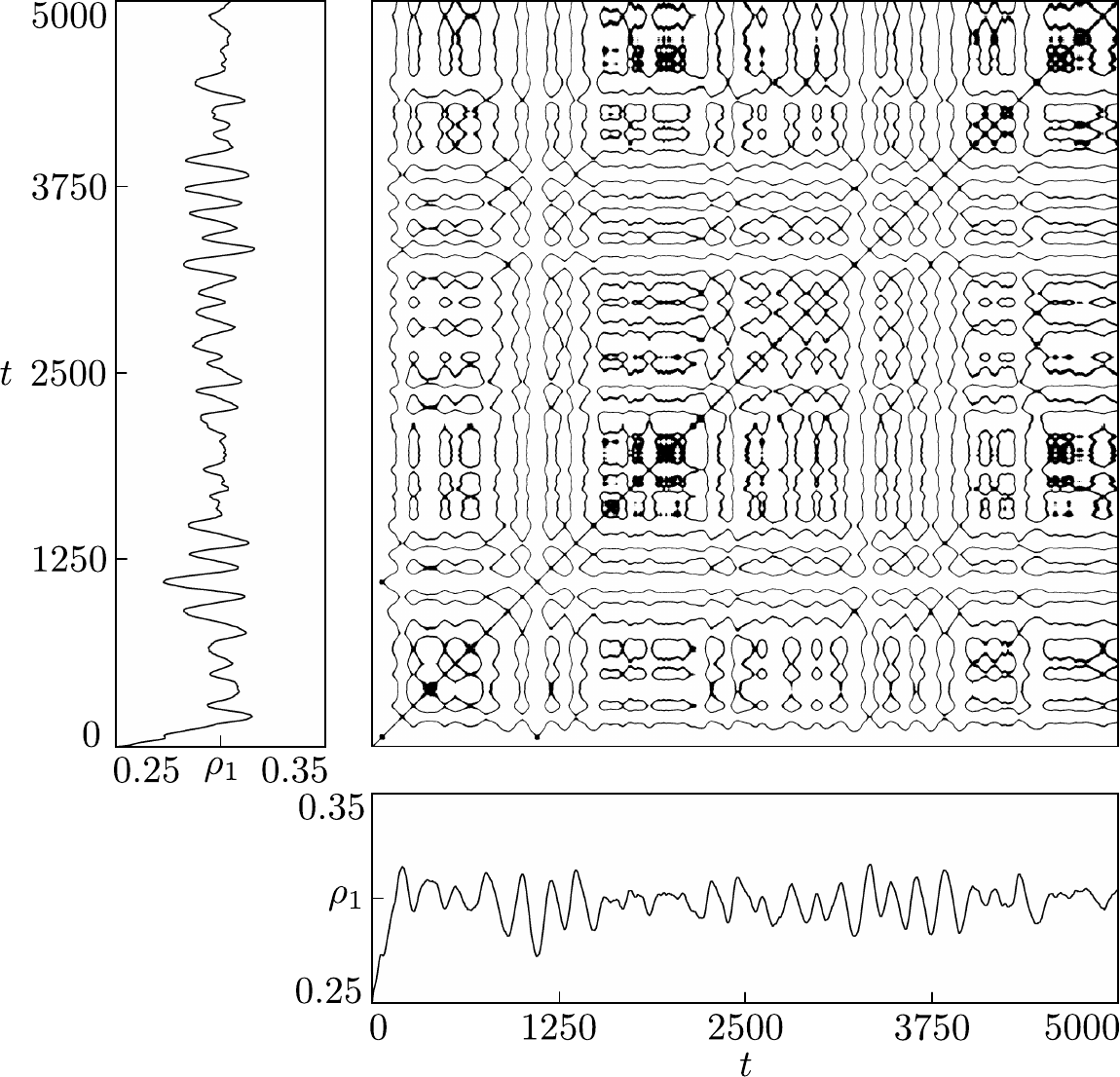}
    \caption{Example of a recurrence plot (centre) constructed with a fixed recurrence rate of $\text{RR} = 5\%$, based on the abundance time-series of species~1 with mobility $m = 0.6$. The simulation was run on a lattice of size $N = 10^3$ and up to $5{,}000$ iterations. The side panels show the corresponding time-series $\rho_1$ used to construct the RP. The two time-series are essentially identical, reflecting the selected minimal delay in the embedding ($\tau = 1$ and $d=2$).}
    \label{fig:example_rp}
\end{figure}

This is accomplished by extracting all distinct off-diagonal elements of $\mathbf{D}$, sorting them into an ordered set,
\begin{equation}
    \{d_1, d_2, \dots, d_M\},~\text{where } M = \frac{N(N-1)}{2},
    \label{eq:sortting_D}
\end{equation}and selecting $\varepsilon$ as the $\lfloor \text{RR} \times M \rfloor$-th smallest distance. Once the optimal threshold $\varepsilon$ is determined, the recurrence matrix is computed as
\begin{equation}
    R_{ij} = \Theta(\varepsilon - D_{ij}),
    \label{eq:RM}
\end{equation}ensuring the predefined RR for all constructions. As the distance matrix $\mathbf{D}$ is symmetric under most norms, the resulting RM is also symmetric. The RP is then visualised by mapping $R_{ij} = 1$ to coloured pixels, in our study to black pixels, and $R_{ij} = 0$ to white pixels. An example of such a plot is shown in Fig.~\ref{fig:example_rp}.

From this point onward, all RPs are constructed using fixed embedding parameters ($\tau = 1$, $d = 2$) and presented in the standard format, displaying only the central panel as in Fig.~\ref{fig:example_rp}. The associated time-series plots are omitted for brevity and visual clarity, focusing instead on the recurrence structures themselves, which form the basis of both qualitative and quantitative analyses in the following sections.

\subsubsection*{Recurrence Quantifiers}

As previously mentioned, the RP in our context provides a graphical representation of the abundance time-series from the RPS model under investigation. However, to extract meaningful insights beyond qualitative inspection, quantitative measures are necessary. Recurrence quantification analysis (RQA) \cite{zbilut92,webber94,marwan2002herz,marwan2007recurrence, MarwanWebber2015} introduces statistical measures that quantify these patterns, characterising the complexity and predictability of the system. These measures rely on the probability distributions of diagonal and vertical line structures within the RP, providing a robust framework for analysing dynamical properties.

For conciseness, we omit the explicit dependence on $\varepsilon$ in the following definitions. The frequency distribution for diagonal lines of length $l$ is given by

\begin{equation}
    P(l) = \sum_{i,j=1}^{N} (1 - R_{i-1,j-1}) (1 - R_{i+l,j+l}) \prod_{k=0}^{l-1} R_{i+k,j+k}.
    \label{eq:p(l)}
\end{equation}

Similarly, the frequency distribution for vertical lines of length $v$ is 
\begin{equation}
    P(v) = \sum_{i,j=1}^{N} (1 - R_{i,j-1}) (1 - R_{i,j+v}) \prod_{k=0}^{v-1} R_{i,j+k}.
    \label{eq:p(v)}
\end{equation}

Using these frequency distributions, the main recurrence quantifiers are defined as follows:

\begin{itemize}
    \item \textbf{Determinism (DET)}: The fraction of recurrence points forming diagonal lines of at least length $l_{\min}$ (typically $l_{\min} = 2$), indicative of deterministic behaviour
    \begin{equation}
        \text{DET} = \frac{\sum_{l=l_{\min}}^{N} l P(l)}{\sum_{l=1}^{N} l P(l)}.
        \label{eq:det}
    \end{equation}
    
    \item \textbf{Laminarity (LAM)}: The fraction of recurrence points forming vertical structures of at least length $v_{\min}$ (typically $v_{\min} = 2$), measuring laminar (intermittent) behaviour
    \begin{equation}
        \text{LAM} = \frac{\sum_{v=v_{\min}}^{N} v P(v)}{\sum_{v=1}^{N} v P(v)}.
        \label{eq:lam}
    \end{equation}
    
    \item \textbf{Divergence (DIV)}: The inverse of the longest diagonal line length $L_{\max}$, providing an estimate of system instability
    \begin{equation}
        \text{DIV} = \frac{1}{L_{\max}} = \frac{1}{\max\{l_i\}_{i=1}^{N_l}},
            \label{eq:div}
    \end{equation}
    where $N_l = \sum_{l>l_{\min}} P(l)$ is the total number of diagonal lines.
    
    \item \textbf{Trapping Time (TT)}: The average length of vertical structures, representing the mean duration of laminar states
    \begin{equation}
        \text{TT} = \frac{\sum_{v=v_{\min}}^{N} v P(v)}{\sum_{v=v_{\min}}^{N} P(v)}.
        \label{eq:tt}
    \end{equation}

    \item \textbf{Diagonal Entropy (ENTR-L)}: The Shannon entropy of the probability distribution of diagonal line lengths, $p(l) = P(l) / \sum_l(P(l))$, providing insight into the complexity of recurrence structures
    \begin{equation}
        \text{ENTR-L} = - \sum_{l=l_{\min}}^{N} p(l) \ln p(l).
            \label{eq:entr-l}
    \end{equation}
    
    \item \textbf{Vertical Entropy (ENTR-V)}: The Shannon entropy of the probability distribution of vertical line lengths, $p(v) = P(v) / \sum_l(P(v))$, characterizing the variability in laminar states
    \begin{equation}
        \text{ENTR-V} = - \sum_{v=v_{\min}}^{N} p(v) \ln p(v).
            \label{eq:entr-v}
    \end{equation}
    
\end{itemize}

These quantifiers allow for a detailed characterization of the different dynamical behaviours that can emerge in the RPS model. In particular, they are capable of detecting fine changes within the complex oscillatory dynamics, as depicted in Fig.~\ref{fig:dst}, enabling a deeper understanding of the system's underlying dynamics.

\subsubsection*{Ensemble Analysis}
In addition to the usual RQA, we propose a methodological approach that leverages ensemble recurrence analysis \cite{Palmero2023, braun2023} to identify variations in the dynamics of the RPS system. The central idea is to monitor the behaviour of a given recurrence quantifier $Q$ (e.g., DET, LAM, \ldots) across an ensemble of simulations, each indexed by an integer $k$. Each index $k$ corresponds to a distinct realisation of the RPS dynamics, producing a unique RP associated with the abundance time-series of a selected species.

By running a large number of independent simulations, we construct a distribution of values $Q(k)$. From this distribution, we compute the mean $\langle Q \rangle$ and the standard deviation $\sigma$, which allow us to define a statistical threshold for identifying atypical behaviour. Realisations for which $Q(k)$ deviates from the mean, i.e., those satisfying

\begin{equation}
\{k^*\} = \left\{ k \, \middle|\ \left|Q(k) - \langle Q \rangle\right| > 3\sigma \right\},
\end{equation}are classified as \emph{outliers}. Within this set, we define

\begin{equation}
k^+ = \arg\max_{k \in \{k^*\}} Q(k) \quad \text{and} \quad k^- = \arg\min_{k \in \{k^*\}} Q(k),
\end{equation}as the realisations corresponding to the most extreme values of the quantifier above and below the threshold, respectively. In what follows, we restrict our attention to these extreme cases, $k^+$ and $k^-$, which serve as representative realisations of pronounced dynamical deviations.

\begin{figure}[h!]
    \centering
    \includegraphics[scale=0.8]{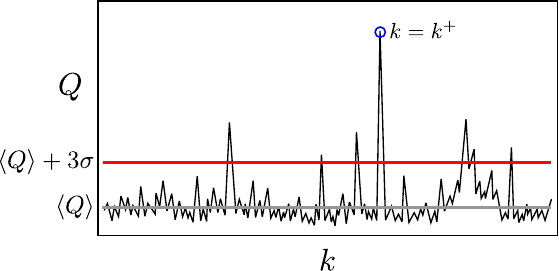}
        \caption{Schematic representation of the ensemble recurrence analysis. Each realisation $k$ yields a recurrence quantifier value $Q(k)$. The ensemble average $\langle Q \rangle$ (grey line) and standard deviation $\sigma$ define a statistical threshold at $\langle Q \rangle + 3\sigma$ (red line). Realisations for which $Q(k) > \langle Q \rangle + 3\sigma$ are identified as $k^*$. Among these, the realisation $k=k^+$ with the maximum quantifier value $Q(k^+)$, is highlighted (blue circle) as the most extreme case.} 
    \label{fig:scheme_rec_ensemble}
\end{figure}

These outlier realisations, specifically $k^+$ and $k^-$, may reveal distinct dynamical patterns within the ensemble and are particularly valuable for identifying anomalies within the ensemble. The overall procedure is illustrated schematically in Fig.~\ref{fig:scheme_rec_ensemble}, where we exemplify considering only the \( \langle Q \rangle + 3\sigma \) threshold.

\begin{figure*}[t!]
    \centering
    \includegraphics[width=16.8cm]{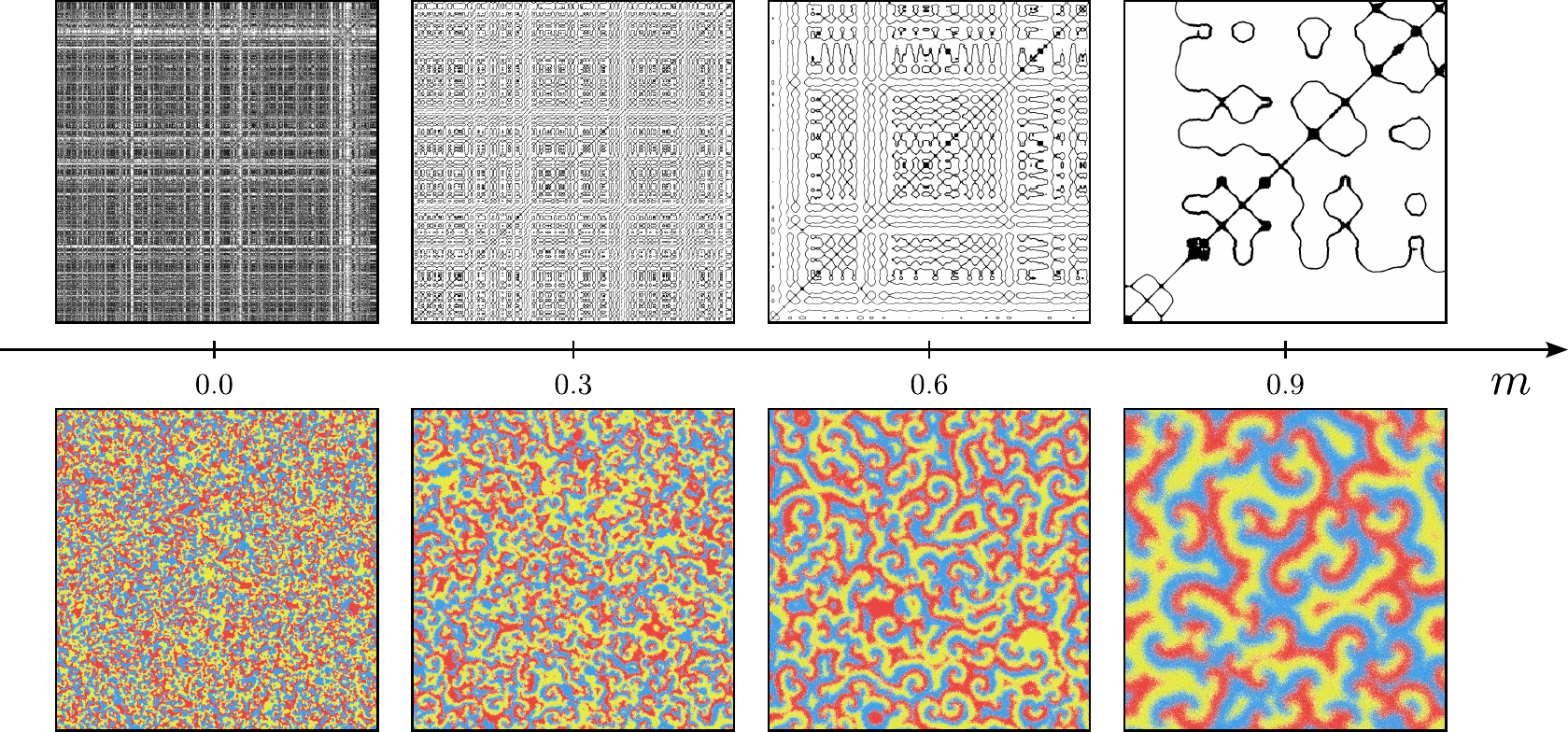}
    \caption{(Upper panels) Recurrence plots constructed from the abundance time-series of species~1 $(\rho_1)$, considering the model simulation with four increasing values of the mobility parameter $m$. (Lower panels) Final states of the lattice simulation considering its size $N=10^3$, and evolved up to $5,000$ generations for the same four values of $m$. White pixels represent empty spaces while red, blue, and yellow pixels represent species $1, 2,$ and $3$, respectively.} 
    \label{fig:panel_snaps_rp}
\end{figure*}

\section{Results}
\label{sec:results}

This section discusses the results of various simulations of the RPS model conducted under the recurrence-based methodology outlined in Sec.~\ref{subsec:methodology}. In the first subsection, we compile the findings from an extensive RQA performed for a range of mobility parameter values $m$ below the extinction threshold. Specifically, we compute the average values of six distinct quantifiers (determinism, laminarity, divergence, trapping time, as well as diagonal and vertical line length entropies) as functions of $m$. Furthermore, in the second subsection, we present a novel approach, based on the recurrence quantifiers, for detecting divergent mobility parameters among an ensemble of many distinct numerical simulations of the model.

It is important to highlight that the qualitative behaviour across all three species in the RPS system is essentially identical, exhibiting analogous complex oscillation and fluctuation patterns. Consequently, for simplicity and clarity, we focus our analysis primarily on the abundance time-series of only the species~1 $(\rho_1)$, represented by the red curve in Fig.~\ref{fig:dst}, and also displayed by the side panels in Fig.~\ref{fig:example_rp}.

As an initial observation and comparison, Fig.~\ref{fig:panel_snaps_rp} presents the final states of the models' dynamics for four different values of the mobility parameter $m$. The snapshot of the final state is a characteristic and representative configuration of the temporal dynamics. In the spatial representation each pixel is coloured by the respective species' individual that occupies a particular site, following the colours defined in Fig.~\ref{fig:rps+rules}. Each state is accompanied by its respective RP, constructed from the simulated abundance curve corresponding to that specific value of $m$. An interesting feature of both the RPs and final states in Fig.~\ref{fig:panel_snaps_rp} is the apparent presence of structured, self-similar patterns, suggesting potential fractal characteristics in the system's dynamics. If these structures indeed exhibit scale invariance, a more detailed fractal analysis, such as the one proposed in \cite{braun2021}, could be insightful. Methods such as box-counting dimension or scaling analysis of recurrence quantifiers could provide further comprehension into the hierarchical nature of the system's recurrences. A comprehensive investigation of these aspects, particularly through scaling analysis techniques, will be pursued in future work.

\subsection{Average Recurrence Quantifiers}
\label{subsec:average_rqa}

To further understand and, more importantly, quantify the dynamical changes in the model, we performed an extensive numerical analysis of six different recurrence quantifiers while gradually increasing the mobility parameter. We avoid values approaching $m \to 1$, as biodiversity is lost in this regime \cite{reichenbach2007mobility}. Accordingly, we explore the interval $m \in [0.00, 0.90]$, with increments of $0.03$.

The results presented below are obtained from an average over 100 independent simulations for each value of $m$, each conducted on a lattice of size $10^3 \times 10^3$. The simulations were evolved up to 5,000 generations, discarding the first 1,000 to eliminate transient effects. The standard error of each measurement is on the order of $10^{-5}$ for all computed quantifiers, remaining within the size of the plotted points. For further details on the computation of standard error, we refer to \cite{altman2005standard}.

\begin{itemize}

    \item \textbf{Determinism:} 

    \begin{figure}[h!]
    \centering
    \includegraphics[width=8.4cm]{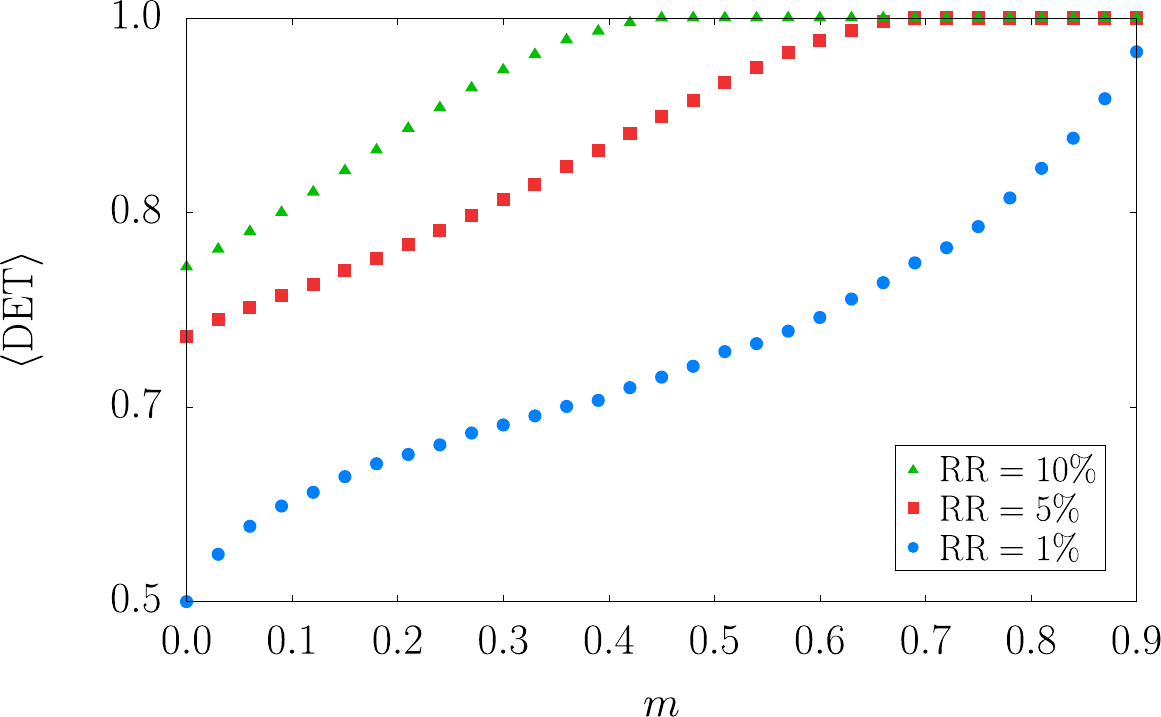}
    \caption{Average DET as a function of the mobility parameter $m$. The analysed RPs were constructed considering three different values of the Recurrence Rate (RR), depicted by the three different set of points.} 
    \label{fig:det}
    \end{figure}
    
    Figure~\ref{fig:det} shows the behaviour of the average determinism $\langle \mathrm{DET} \rangle$ as a function of the mobility parameter $m$, evaluated for three different recurrence rates: RR = 1\%, 5\%, and 10\%. In all cases, we observe a monotonic increase in $\langle \mathrm{DET} \rangle$ with $m$, indicating a progressive enhancement in the temporal regularity of the system's dynamics as the formation of the dynamical spiral patterns become more and more present. For higher RR (5\% and 10\%), $\langle \mathrm{DET} \rangle$ quickly saturates near unity for intermediate values of $m$, suggesting the emergence of strongly deterministic behaviour. In contrast, for RR = 1\%, the growth of $\langle \mathrm{DET} \rangle$ is more gradual and does not fully saturate within the explored range. This implies that lower RR may capture subtler dynamical features, preserving fluctuations that are otherwise smoothed out at higher RR values. 

    \item \textbf{Laminarity:} 

    \begin{figure}[h!]
    \centering
    \includegraphics[width=8.4cm]{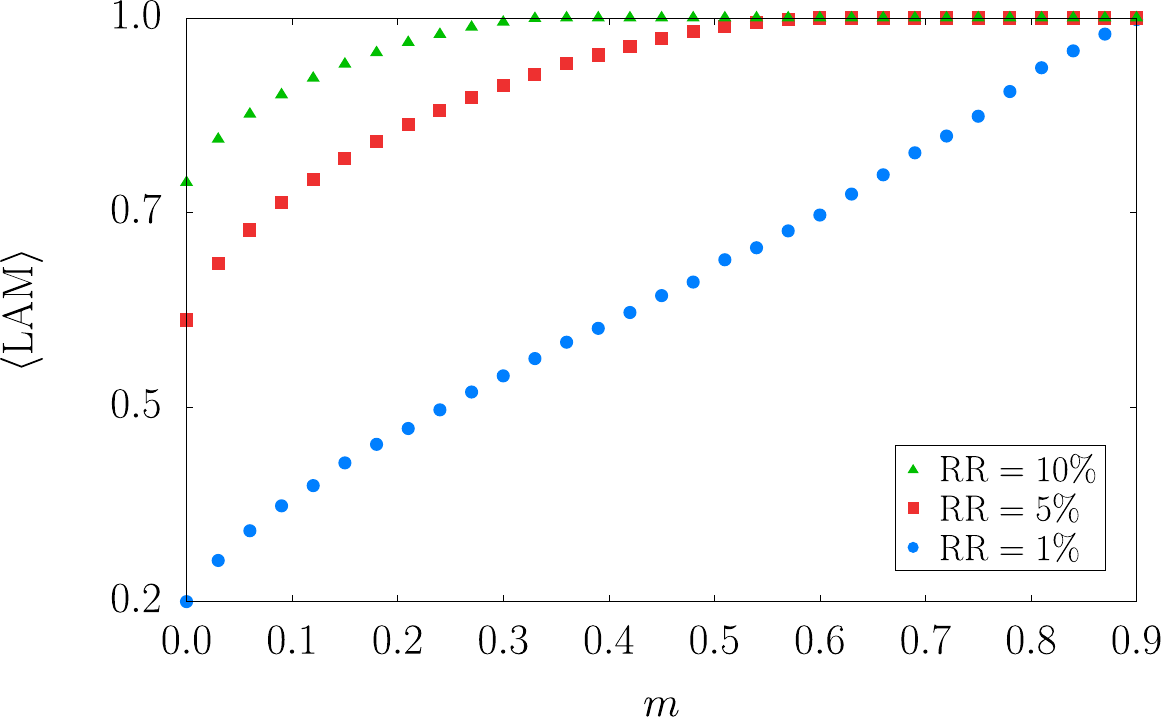}
    \caption{Average LAM as a function of the mobility parameter $m$. The analysed RPs were constructed considering three different values of RR, depicted by the different sets of points.}
    \label{fig:lam}
    \end{figure}

    As with determinism, we observe a monotonic increase in $\langle \mathrm{LAM} \rangle$ across all RR values, indicating a growing prevalence of laminar phases in the system's dynamics (Fig.~\ref{fig:lam}). For RR = 5\% and 10\%, the curves rapidly saturate near unity, while for RR = 1\% the increase is more gradual, capturing finer temporal fluctuations. The lowest values of $\langle \mathrm{LAM} \rangle$ are observed at $m \approx 0.00$, particularly for RR = 1\%, suggesting a more turbulent-like behaviour at low mobility. These results support the interpretation that the system transitions from a turbulent to a laminar-like regime as $m$ increases, likely reflecting the emergence of coherent wave fronts.

    \item \textbf{Trapping Time:} 

    \begin{figure}[h!]
    \centering
    \includegraphics[width=8.4cm]{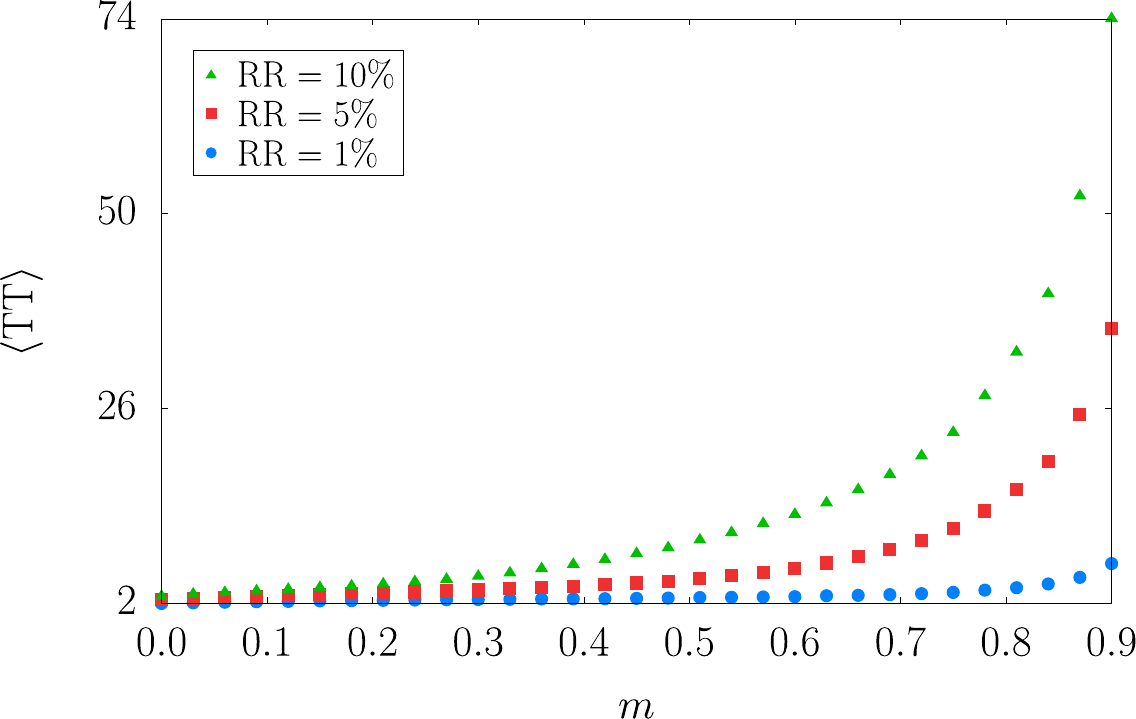}
    \caption{Average TT as a function of the mobility parameter $m$. The analysed RPs were constructed considering three different values of RR, depicted by the different sets of points.}
    \label{fig:tt}
    \end{figure}

    In contrast to the monotonic behaviour observed in the previous measures, $\langle \mathrm{TT} \rangle$ shows a markedly non-linear growth as the mobility parameter $m$ increases (Fig.~\ref{fig:tt}). For low values of $m$, $\langle \mathrm{TT} \rangle$ remains nearly constant and close to its minimal value, indicating short laminar phases and frequent switching in the dynamics. As $m$ exceeds approximately 0.4, the quantifier begins to rise more sharply, especially for higher RR, highlighting the onset of longer laminar intervals and more persistent dynamical states. This steep growth at larger mobility values reflects a dynamical slowing down, in which the system becomes increasingly trapped in low-variability episodes, consistent with the formation of coherent wave structures due to the presence of the spiral structures. Assuming an underlying exponential dependence in this regime, such behaviour indicates scaling features in the trapping dynamics, further suggesting the presence of fractal or self-similar structures, as discussed earlier in this section.

    \item \textbf{Divergence:} 

    \begin{figure}[h!]
    \centering
    \includegraphics[width=8.4cm]{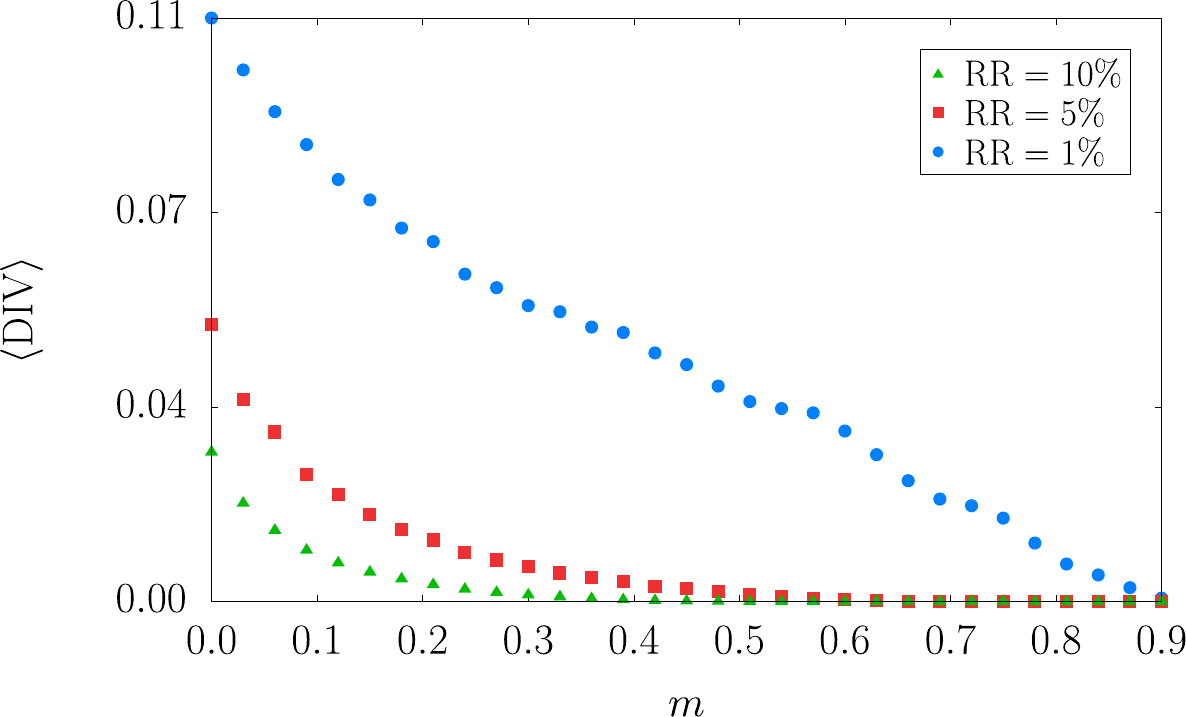}
    \caption{Average DIV as a function of the mobility parameter $m$. The analysed RPs were constructed considering three different values of RR, depicted by the different sets of points.}
    \label{fig:div}
    \end{figure}

    In contrast to the increasing trends observed for the previous quantifiers, the divergence exhibits a rapid decay as $m$ increases, approaching zero for all RR beyond $m \approx 0.5$ (Fig.~\ref{fig:div}). Since $\langle \mathrm{DIV} \rangle$ is inversely related to the length of the longest diagonal line, Eq.(\ref{eq:div}), this result reflects a decreasing level of dynamical instability at higher mobility values. The quantifier is largest around $m \approx 0.00$, particularly for lower RR, suggesting more chaotic-like dynamics in the low-mobility regime as individuals predate one another more frequently. As $m$ increases, the system becomes progressively more coherent and predictable, consistent with the transition toward laminar and wave-like dynamics highlighted by other recurrence measures.

    \item \textbf{Diagonal and Vertical Lines Length Entropy:} 

    \begin{figure}[h!]
    \centering
    \includegraphics[width=8.4cm]{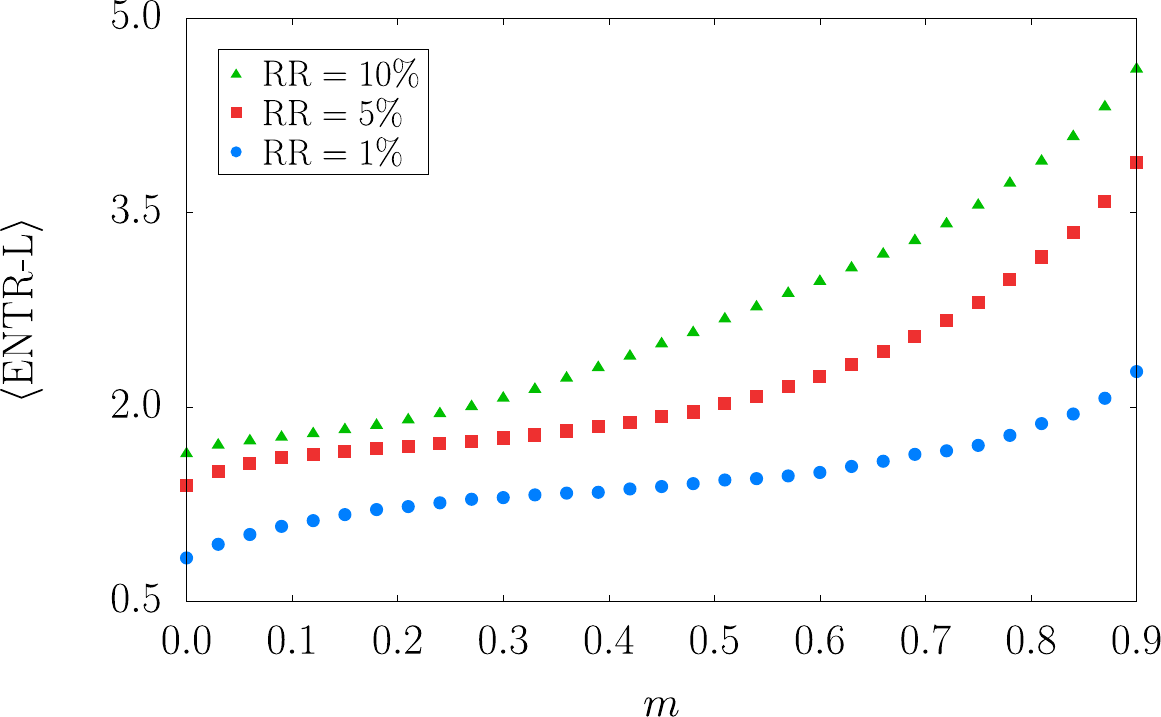}
    \caption{Average ENTR-L as a function of the parameter $m$. The analysed RPs were constructed considering three different values of RR, depicted by the different sets of points.}
    \label{fig:entr-l}
    \end{figure}

    \begin{figure}[h!]
    \centering
    \includegraphics[width=8.4cm]{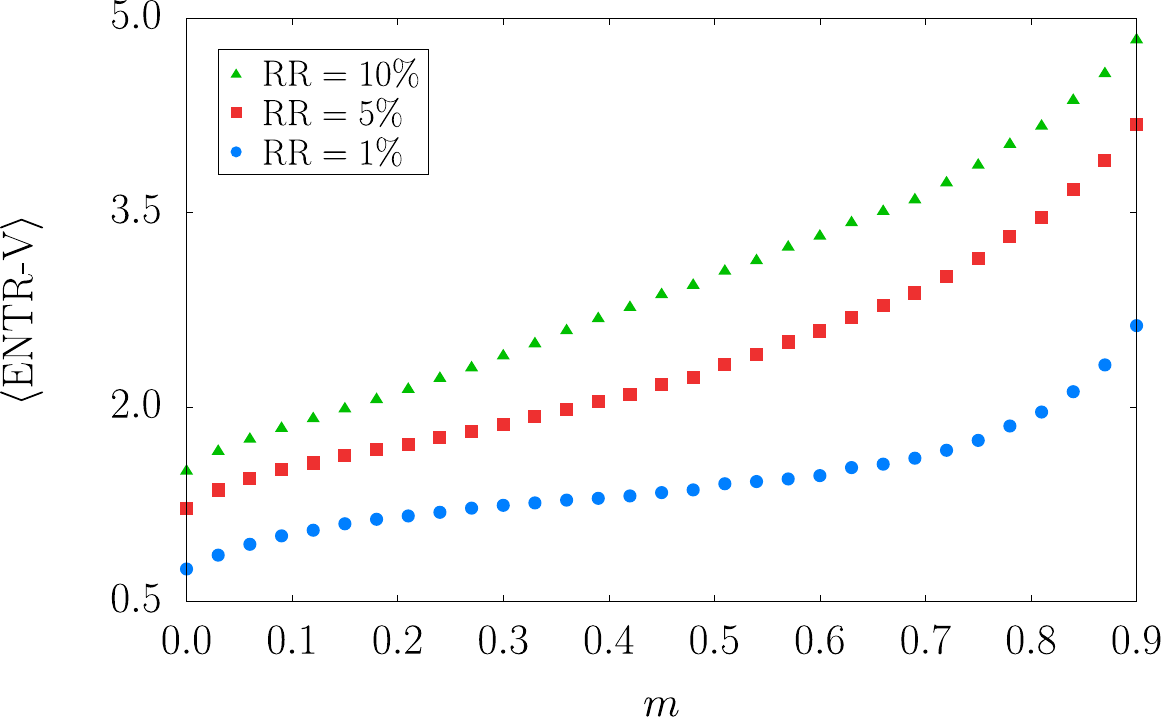}
    \caption{Average ENTR-V as a function of the parameter $m$. The analysed RPs were constructed considering three different values of RR, depicted by the different sets of points.}
    \label{fig:entr-v}
    \end{figure}

    Both quantifiers exhibit a generally smooth and accelerating increase with $m$, particularly for higher RR values. This trend indicates an increase in the diversity of diagonal and vertical line lengths within the RPs, which in turn reflects richer dynamical variability as mobility increases.

    At first glance, this result might seem counter-intuitive. As illustrated in Fig.~\ref{fig:panel_snaps_rp}, higher mobility values appear to yield larger and more coherent spatio-temporal patterns within the RP, which could initially suggest more regularity. However, it is crucial to understand that the entropy measures computed here are not derived from the regularity or visual coherence of patterns themselves, but rather from the statistical distribution of line lengths in the RPs. Therefore, even visually structured recurrence patterns can contain highly variable recurrence intervals and laminar durations, thus increasing the entropy.

    Moreover, a detailed examination of these recurrence structures reveals an additional subtlety: the irregular and continuously evolving shape of dynamical cycles contributes to variability not only in the apparent line lengths visible in the RPs but also in the thickness of these lines. This line-thickness variability represents an essential feature that is captured by the line length distribution, which consequently increases the entropy measures. The line-thickness variability is specially significant for high mobility values, as depicted in the last panel of Fig.\ref{fig:panel_snaps_rp}, and become even more evident in the analysis presented in Fig.\ref{fig:rps_10}.

    Finally, it is important to stress that all quantifier values reported here were obtained using recurrence plots constructed with consistent embedding parameters and computational settings. Although alternative techniques, such as skeletonisation~\cite{kraemer2019} or changes in threshold criteria~\cite{marwan2007recurrence}, could affect the absolute values of these quantifiers, our focus is specifically on how they vary systematically with increasing mobility $m$, maintaining a controlled comparative framework.

\end{itemize}

\subsection{Mobility Detection}
\label{subsec:mobility_detect}

\begin{figure}[h!]
    \centering
    \includegraphics[width=8.4cm]{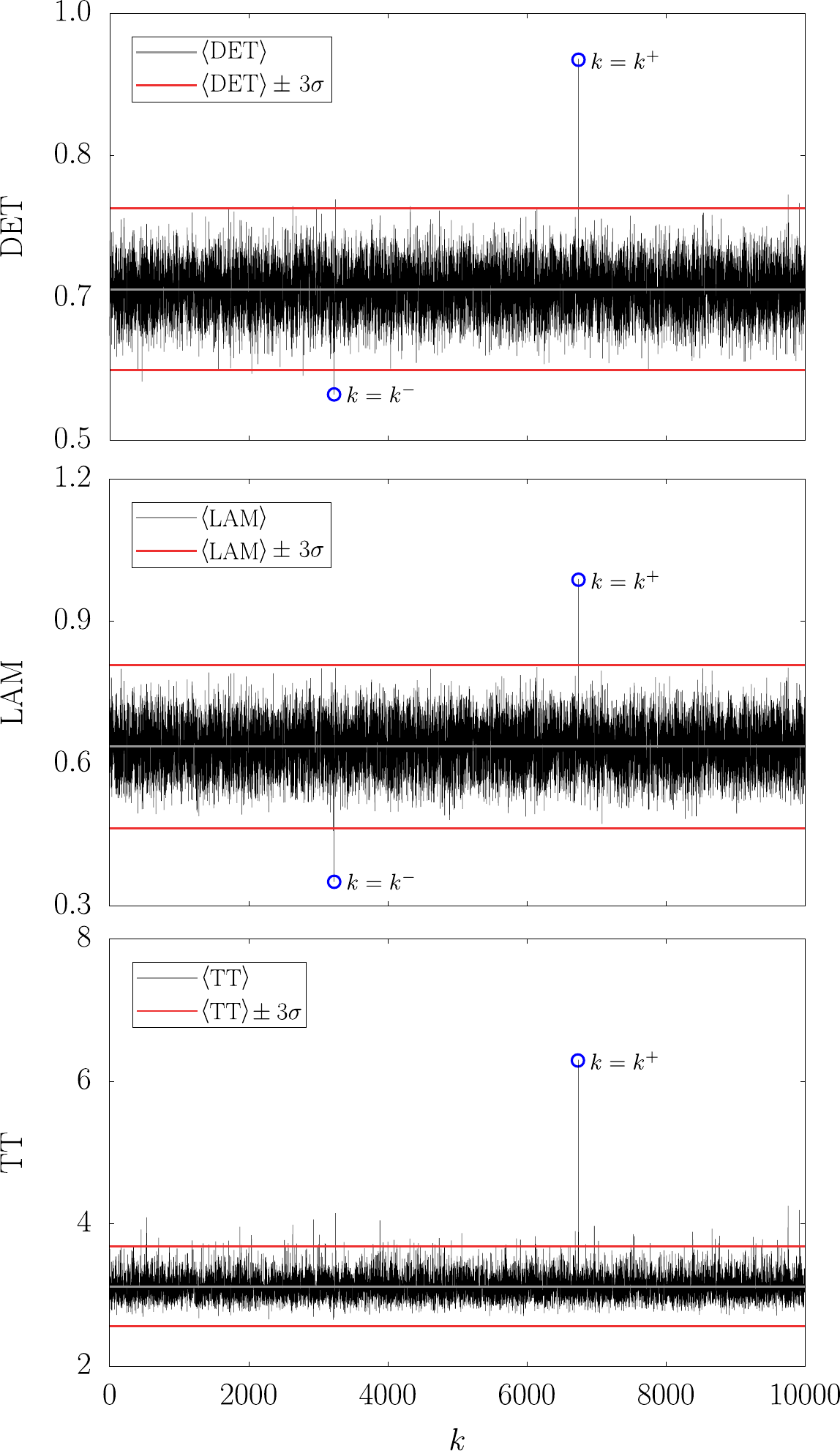}
    \caption{Distributions of determinism (DET), laminarity (LAM), and trapping time (TT) across an ensemble of $10^4$ distinct simulations of the RPS model. These quantifiers were computed from RPs constructed with a fixed percentage $RR = 1\%$. Outlier simulations, detected via the statistical threshold $ \langle Q \rangle \pm 3\sigma$ and with the largest quantifier deviations from the ensemble mean, are highlighted by the blue circles.} 
    \label{fig:dist_1}
\end{figure}

\begin{figure}[h!]
    \centering
    \includegraphics[width=8.66cm]{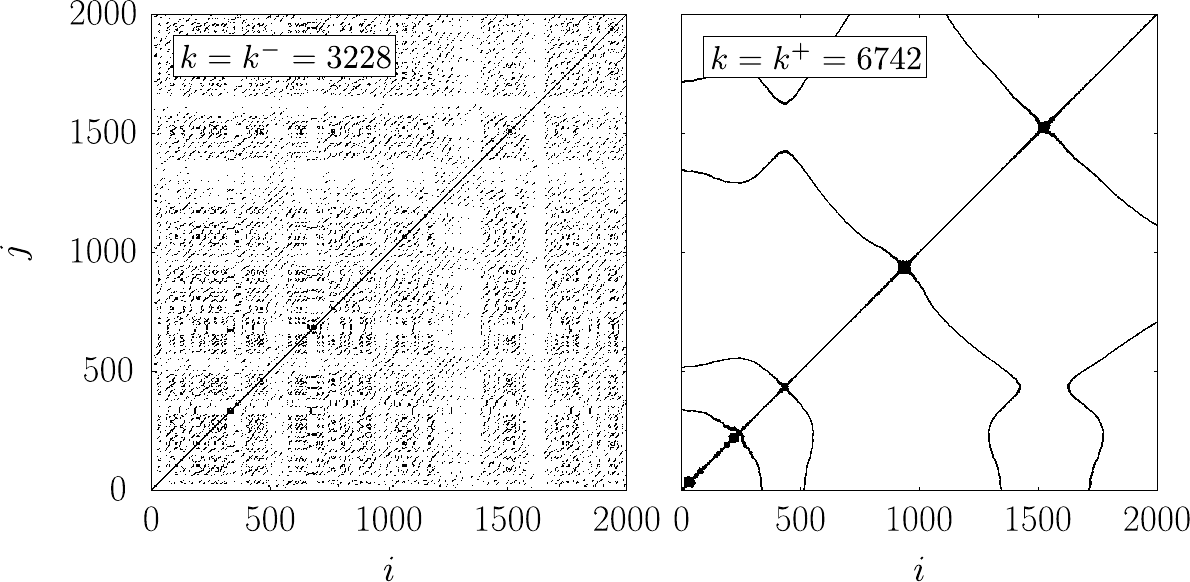}
    \caption{Recurrence plots constructed with $RR = 1\%$ and associated with the two outlier simulations identified in Fig.~\ref{fig:dist_1}, located at $k = k^- = 3228$ (left) and $k = k^+ = 6742$ (right). These simulations correspond to particularly low and high values of the mobility parameter $m$, respectively.} 
    \label{fig:rps_1}
\end{figure}

\begin{figure}[h!]
    \centering
    \includegraphics[width=8.4cm]{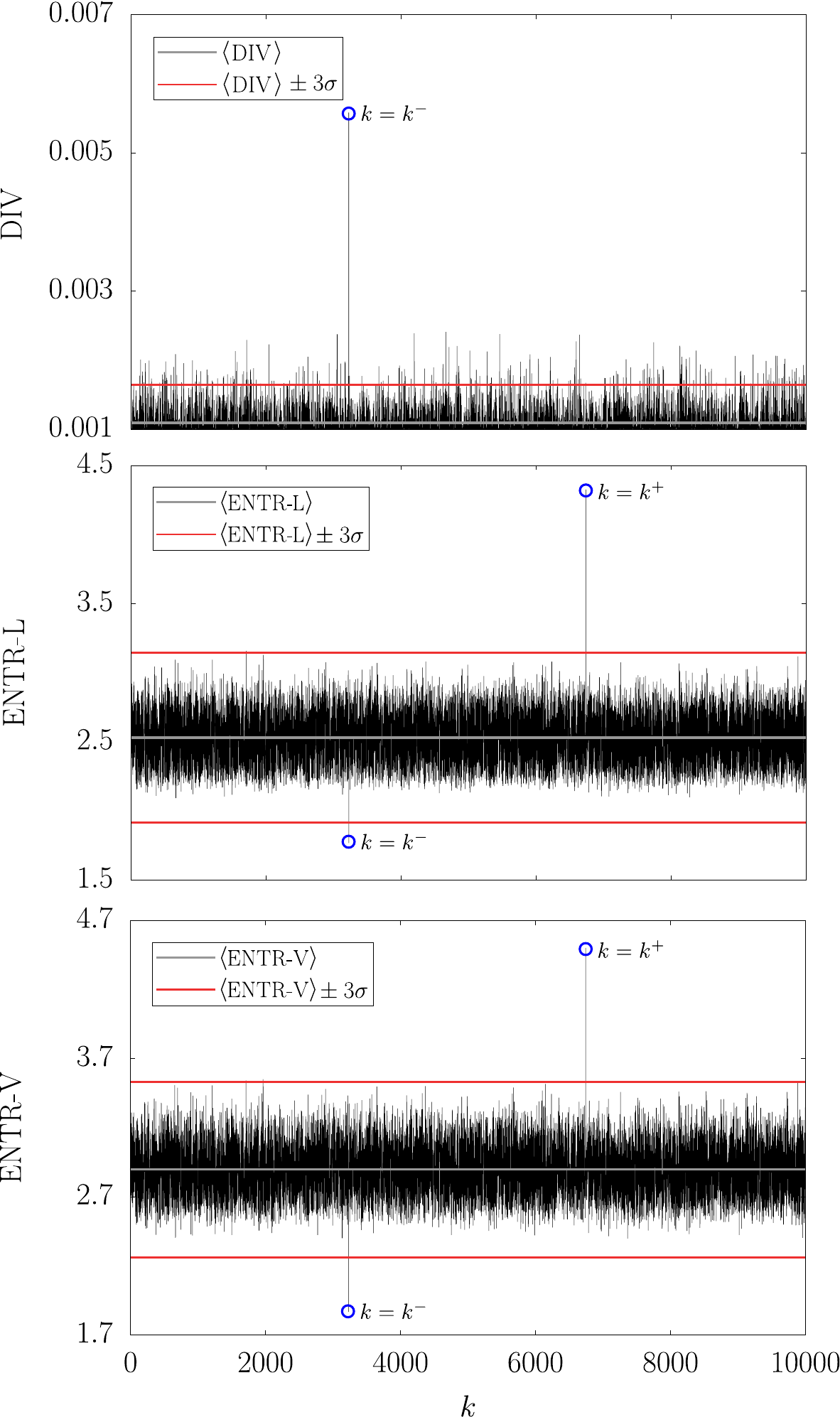}
    \caption{Distributions of Divergence (DIV), Diagonal Entropy (ENTR-L), and Vertical Entropy (ENTR-V) across an ensemble of $10^4$ distinct simulations of the RPS model. These quantifiers were computed with a fixed percentage $RR = 10\%$. Outlier simulations, detected via \( \langle Q \rangle \pm 3\sigma \) and with the largest quantifier deviations from the ensemble mean, are highlighted with blue circles.} 
    \label{fig:dist_10}
\end{figure}

\begin{figure}[h!]
    \centering
    \includegraphics[width=8.66cm]{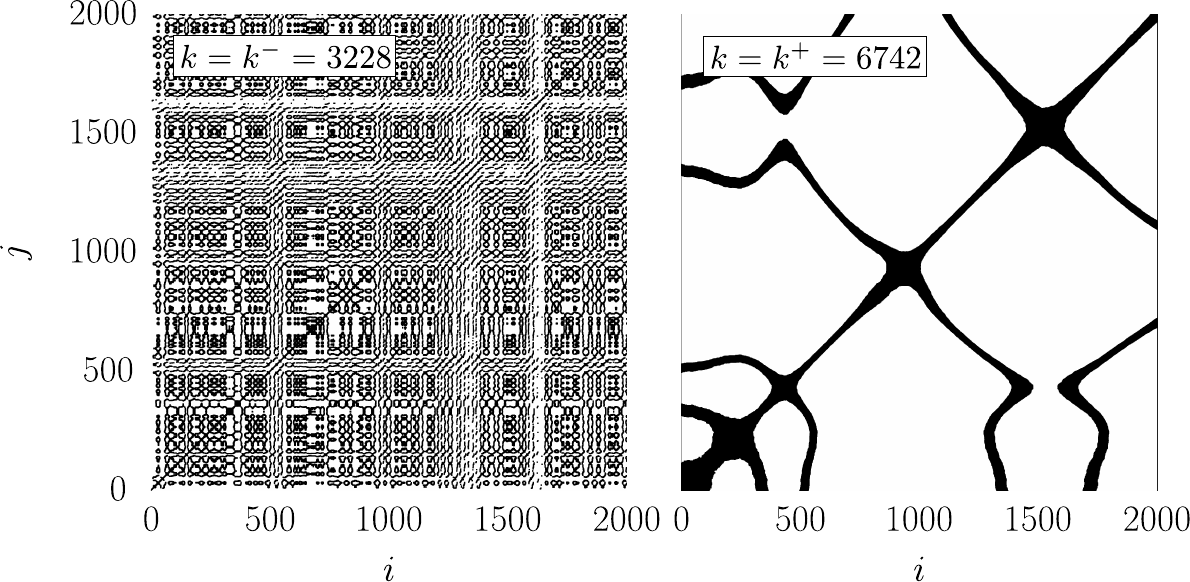}
    \caption{Recurrence Plots constructed with $RR = 10\%$ for the two outlier simulations identified in Fig.~\ref{fig:dist_10}, located at $k^- = 3228$ (left) and $k^+ = 6742$ (right).} 
    \label{fig:rps_10}
\end{figure}

Based on the ensemble analysis described at the end of Sec.~\ref{subsec:methodology}, we consider an additional set of $10^4$ independent simulations of the RPS model as our ensemble. Each simulation is performed on a fixed lattice of size $N = 10^3$ and evolved for 3{,}000 generations, with the initial 1{,}000 discarded to exclude transient effects. Thus, the effective length of each time-series is $T = 2000$, which also determines the dimensions of all resulting RPs. The mobility parameter $m$ is randomly sampled from the interval $m \in [0.35,\ 0.55]$, introducing controlled variability into the system's dynamics. Additionally, two simulations corresponding specifically to particularly low and high mobility values are included and inserted at random positions $k$ within the ensemble. These two cases allow us to assess the sensitivity of the various recurrence quantifiers $Q$ in detecting dynamical deviations across the ensemble.

Initially, with the recurrence percentage fixed at $\text{RR} = 1\%$, we computed the recurrence quantifiers determinism (DET), laminarity (LAM), and trapping time (TT) across all $10^4$ RPS simulations. Figure~\ref{fig:dist_1} presents the resulting distributions. We adopt $\text{RR} = 1\%$ for DET, LAM, and TT as it yields the most linear behaviour across the quantifiers, as evident in Figs.\ \ref{fig:det}, \ref{fig:lam}, and \ref{fig:tt}. As previously established, simulations whose quantifier values deviate beyond the statistical threshold of $\langle Q \rangle \pm 3\sigma$ are identified as outliers, and we are interested in the largest deviations. These outliers, exemplified at positions $k = k^{-}$ and $k = k^{+}$ (highlighted by the blue circles), represent simulations exhibiting distinct dynamical regimes, potentially associated with contrasting values of the mobility parameter $m$.

To further investigate the dynamical differences identified through mobility detection, Fig.~\ref{fig:rps_1} presents the RPs corresponding to the outlier simulations indicated in Fig.~\ref{fig:dist_1}, specifically at $k = k^- = 3228$ (left panel) and $k = k^+ = 6742$ (right panel). These cases represent the simulations with the largest quantifier deviations from the ensemble mean.

The recurrence plot on the left, corresponding to $k^-$, exhibits a fragmented structure characterised by numerous short diagonal and vertical lines, indicative of intermittent and less predictable dynamics. This pattern is typical of chaotic or irregular behaviour, consistent with lower mobility. In contrast, the recurrence plot on the right, corresponding to $k^+$, displays a markedly regular structure, distinguished by long uninterrupted diagonal lines and clearly defined recurrence domains. Such a pattern suggests stable or periodic dynamics, typically arising from higher mobility values.

This visual comparison supports the interpretation that mobility strongly influences the temporal recurrence patterns in the RPS system, demonstrating that the proposed ensemble recurrence methodology effectively differentiates these dynamical regimes through straightforward statistical analysis of recurrence quantifiers.

Following the same procedure, Fig.~\ref{fig:dist_10} shows the distributions of the divergence (DIV), diagonal line length entropy (ENTR-L), and vertical line length entropy (ENTR-V), now computed from recurrence plots constructed with a higher recurrence percentage, $\text{RR} = 10\%$. Again, we adopted $\text{RR} = 10\%$ in this case as it yields the most linear behaviour, as evident in Figs.\ \ref{fig:div}, \ref{fig:entr-l}, and \ref{fig:entr-v}. Since the analysis is performed on the same dataset as in Fig.~\ref{fig:dist_1}, the two special simulations with particularly low and high mobility remain at the same positions $k = k^-$ and $k = k^+$. These simulations are again identified as statistical outliers (highlighted by the blue circles), confirming that these quantifiers are also sensitive to dynamical deviations induced by variations in mobility.

As previously observed in Fig.~\ref{fig:rps_1}, the RPs corresponding to the outlier simulations at $k = k^- = 3228$ (left) and $k = k^+ = 6742$ (right) exhibit markedly distinct structures, reflecting their underlying dynamical differences. Fig.~\ref{fig:rps_10} displays these RPs using a higher recurrence rate of $RR = 10\%$. The visual patterns remain consistent with those seen at $RR = 1\%$: the RP for $k^-$ shows a highly fragmented and complex structure, whereas the RP for $k^+$ exhibits smooth, periodic patterns with prominent diagonal lines. Owing to the increased RR, these features now appear bolder and more clearly defined. This result further corroborates the robustness of the proposed methodology, demonstrating its effectiveness in detecting mobility across different recurrence rate settings.

\begin{table}[h]
\caption{Values of the recurrence quantifiers for the outlier simulations identified at $k = k^-$ and $k = k^+$. Values in boldface are the ones used for the inference of the value of the mobility parameter $m$.}\label{tab:mobility_quantifiers}
    \begin{tabular}{c|ccc|ccc}
    & \multicolumn{3}{c|}{RR = 1\%} & \multicolumn{3}{c}{RR = 10\%} \\
    $k$ & DET & {\bf LAM} & TT & DIV & ENTR\text{-}L & {\bf ENTR\text{-}V} \\
    \hline
    $k^{-} = 3228$ & 0.586 & {\bf 0.369} & --    & 0.005 & 1.768 & {\bf 1.868} \\
    $k^{+} = 6742$ & 0.950 & {\bf 0.994} & 6.305 & --    & 4.326 & {\bf 4.500} \\
    \end{tabular}
\end{table}

\begin{table}
\end{table}

To quantify the dynamics of these outlier simulations, Tab.~\ref{tab:mobility_quantifiers} presents the values of each recurrence quantifier, computed at their respective recurrence percentages, for the detected cases $k^- = 3228$ and $k^+ = 6742$. By combining these values with the average trends discussed in Sec.~\ref{subsec:average_rqa}, we can infer the mobility parameter values $m(k^-)$ and $m(k^+)$ for these simulations. To that end, we analyse Figs.~\ref{fig:det} to \ref{fig:entr-v} and identify quantifiers that could be approximated with a linear model. For $RR = 1\%$, the average laminarity (blue points in Fig.~\ref{fig:lam}) and for $RR = 10\%$, the average vertical line length entropy (green points in Fig.~\ref{fig:entr-v}) are selected for this purpose. Although the underlying trends are not perfectly linear, a simple linear regression can serve as a first-order approximation to estimate the mobility $m$ from the observed values of LAM or ENTR-V. The fitted coefficients are summarised in Table~\ref{tab:regression_coefficients}.

\begin{table}[h]
\centering
\caption{Values of the linear regression model estimating recurrence quantifiers as a function of mobility $m$. It follows $Q(m) = \beta_1 m + \beta_0$, and $R^2$ is the coefficient of determination (goodness-of-fit). Uncertainties were obtained through weighted linear regression, considering the propagation of the standard errors of $\langle{Q}(m)\rangle$, Figs.\ \ref{fig:lam} and \ref{fig:entr-v}.}
\label{tab:regression_coefficients}
\begin{tabular}{lccc}
\toprule
$Q$ & $\beta_0$ (intercept) & $\beta_1$ (slope) & $R^2$ \\
\midrule
$\mathrm{LAM}$           & $0.247 \pm 0.005$ & $0.834 \pm 0.009$ & 0.996 \\
$\mathrm{ENTR\text{-}V}$ & $1.518 \pm 0.017$ & $3.057 \pm 0.055$ & 0.983 \\
\bottomrule
\end{tabular}
\end{table}

While more complex polynomial or non-linear models could be applied to other quantifiers, our analysis is limited to those exhibiting monotonic trends that justify a linear approach. In this context, the fitted regressions yield explicit inverse relations to estimate the mobility parameter $m$ from the values of LAM and ENTR-V, given by the following equations:

\begin{align}
m &= \frac {\mathrm{LAM}- 0.247}{0.834} \label{eq:m_from_lam},\\
m &= \frac{\mathrm{ENTR\text{-}V} - 1.518}{3.057} \label{eq:m_from_entrv}.
\end{align}

Using the relationships provided by Eqs.~(\ref{eq:m_from_lam}) and (\ref{eq:m_from_entrv}), combined with the highlighted values presented in Tab.~\ref{tab:mobility_quantifiers}, we independently estimate the mobility parameter from the recurrence quantifiers LAM and ENTR-V as follows: 

\begin{table}[h]
\centering
\caption{Values of the recurrence quantifiers $Q(k)$ for both outliers $k^-$ and $k^+$, along with the corresponding mobility values $m(k^-)$ and $m(k^+)$ inferred from the inverse regression relations. Uncertainties in $m(k)$ were calculated via standard error propagation using the regression parameters and their associated uncertainties.}
\label{tab:mobility_k_outliers}
\begin{tabular}{lcccc}
\toprule
$Q$ & $Q(k^-)$ & $m(k^-)$ & $Q(k^+)$ & $m(k^+)$ \\
\midrule
LAM     & 0.369  & $0.146 \pm 0.005$ & 0.994  & $0.895 \pm 0.006$ \\
ENTR-V  & 1.868  & $0.115 \pm 0.004$ & 4.500  & $0.976 \pm 0.014$ \\
\bottomrule
\end{tabular}
\end{table}

To consolidate these independent estimates, we used inverse-variance weighting, which gives greater importance to estimates with smaller uncertainties. This approach yields a weighted average where each mobility estimate is weighted by the inverse of the square of its standard error. The resulting consolidated mobility inferences are:
\[ 
m(k^-) = 0.127 \pm 0.003
\]
and
\[ 
m(k^+) = 0.908 \pm 0.006.
\]

These inferred mobility values can now be directly compared with the known simulated values: $m_\text{sim}(k^-) = 0.10$ and $m_\text{sim}(k^+) = 0.90$. Although the linear models used may not perfectly capture the underlying non-linear trends in the data, the final agreement between the inferred values and the known simulated values is satisfactory. This confirms that the proposed methodology not only captures the qualitative mobility difference between the outlier cases but also provides quantitatively reasonable estimates of the underlying mobility parameter.

These results, however, are inherently tied to the specific simulation parameters and analysis settings. In particular, the accuracy of the inferred mobility values may depend on the lattice size $N$, the recurrence percentage $RR$, or the time-series length $T$. While the present analysis demonstrates clear success under a fixed configuration, further work is needed to assess the robustness of the procedure. A natural extension of this study would involve systematically testing its sensitivity to these parameters, as well as evaluating its scalability in more complex ensemble scenarios.

Finally, as a way to compare the results obtained via the proposed recurrence-based methodology, Appendix~\ref{appendix_A} presents a complementary analysis based on classical time-series measures. There, we investigate whether simple scalar descriptors, namely variance, volatility, and dominant frequency, are able to capture the same outlier simulations detected through the recurrence quantifiers. This comparison exemplifies the limitations of these traditional measures in identifying subtle dynamical structures across the ensemble.

\section{Conclusions and Perspectives}
\label{sec:conclude}

In this paper, we explored how recurrence quantification analysis can improve our understanding of the role of mobility in spatial biodiversity models, specifically within the cyclic, non-hierarchical May-Leonard system, also known as the RPS model. By systematically varying the mobility parameter $m$ and analysing the resulting recurrence structures, we successfully characterised distinct dynamical regimes of the system. To ensure robustness and statistical reliability, we proposed an ensemble-based approach, computing recurrence quantifiers from a large number of independent numerical simulations. This methodology not only enabled us to examine typical dynamical patterns across the ensemble, but also to identify outlier simulations representing distinct ecological states with clearly differentiated dynamics.

From our numerical analysis, we observed that recurrence quantifiers are highly sensitive to variations in the mobility parameter, capturing both subtle and pronounced changes in the dynamics of the RPS model. As mobility increases, the system undergoes a clear transition from irregular, turbulent-like dynamics to more structured and coherent patterns, as reflected in the rise of determinism and laminarity. Complementary insights were obtained through the recurrence-entropy measures, which indicated a progressive shift toward more predictable and organised ecological regimes. Building on this, we introduced a statistical procedure, termed mobility detection, designed to identify outlier dynamics within large ensembles of simulations. By analysing the statistical distributions of recurrence quantifiers, we were able to detect simulations that exhibited atypical dynamical behaviour. Moreover, by establishing simple linear relations between suitable recurrence quantifiers and the mobility parameter, we showed that it is possible to infer, with good accuracy, the underlying mobility values associated with these outliers. Together, these results confirm the utility of recurrence-based methods not only in distinguishing dynamical regimes, but also in quantitatively estimating ecological parameters.

Our findings hold particular promise for addressing fundamental ecological questions concerning the collapse or maintenance of biodiversity, particularly the inference of key parameters such as species mobility from limited observational data originating from ecological systems that can be reasonably modelled by the rock–paper–scissors framework or, more generally, by May--Leonard-type dynamics. In such systems, direct measurements of mobility are often sparse or incomplete due to practical constraints. The ensemble-based recurrence analysis introduced here provides a framework to identify distinct dynamical regimes and to infer the underlying mobility parameter associated with a given dataset. By systematically relating observed ecological patterns to recurrence quantifiers calibrated through simulations, this approach may help bridge theoretical insights and practical ecological monitoring, offering a potential tool for the early detection of critical transitions toward biodiversity loss.

Compared with conventional scalar time-series descriptors, as shown in Appendix~\ref{appendix_A}, recurrence-based quantifiers proved more effective at identifying subtle dynamical distinctions. Traditional scalar measures failed to consistently detect both of significant outliers, highlighting the advantage of recurrence analysis in capturing the complexity of spatial biodiversity models. Although recurrence-based methods are not commonly applied in the context of spatial ecological systems within theoretical ecology, our results suggest that they hold considerable promise. In light of this, several possible directions for future research can be identified. Among them, we highlight the following:

\begin{itemize}
\item \textbf{Real-life applications:} Investigating recurrence quantification methods on empirical ecological data, particularly from spatially structured ecosystems exhibiting cyclic competitive interactions that can be reasonably modelled by RPS systems;

\item \textbf{Scaling analysis:} Employing recurrence-based fractal and scaling techniques to investigate the presence and nature of hierarchical or self-similar structures, both in spatial patterns and within the recurrence plots themselves. For that, the quantifier recurrence lacunarity \cite{braun2021} would be suitable;

\item \textbf{Spatial recurrence techniques:} Extending the analysis using spatio-temporal recurrence methods to further explore species interactions and collective dynamics in the spatial domain \cite{marwan2007pla,riedl2015,marwan2015}
\end{itemize}

Overall, our results demonstrate the potential of recurrence quantification analysis as an insightful framework for studying and inferring key ecological parameters in spatial biodiversity models. We expect that the integration of recurrence-based methods into ecological theory will offer new tools for examining the rich dynamical complexity of ecological systems and, in doing so, contribute to the further understanding of biodiversity patterns in natural ecosystems.

\section*{Code Availability}
The code developed for the analysis presented in this work is available at \href{https://github.com/m-palmero/recurrences_RPS_models}{RQA} and \href{https://github.com/athemusb/recurrence_RPS_models}{RPS}. 

\begin{acknowledgments}
This work was supported by the São Paulo Research Foundation (FAPESP) grants 2023/07704-5 and 2024/22136-6, and by the Coordena\c{c}\~ao de Aperfei\c{c}oamento de Pessoal de N\'ivel Superior (CAPES) grant 88887.688488/2022-00.
\end{acknowledgments}

\section*{Author Contribution Statement}
\textbf{M. P. S.:} Conceptualization, Investigation, Visualization,  Writing - original draft.
\textbf{M. B.:} Conceptualization, Investigation, Visualization,  Writing - review \& editing - original draft.
\textbf{N. M.:} Supervision, Investigation, Writing - review \& editing.

\appendix 
\section{Comparison to linear time-series analysis}        
\label{appendix_A}

\begin{figure}[h!]
    \centering
    \includegraphics[width=8.4cm]{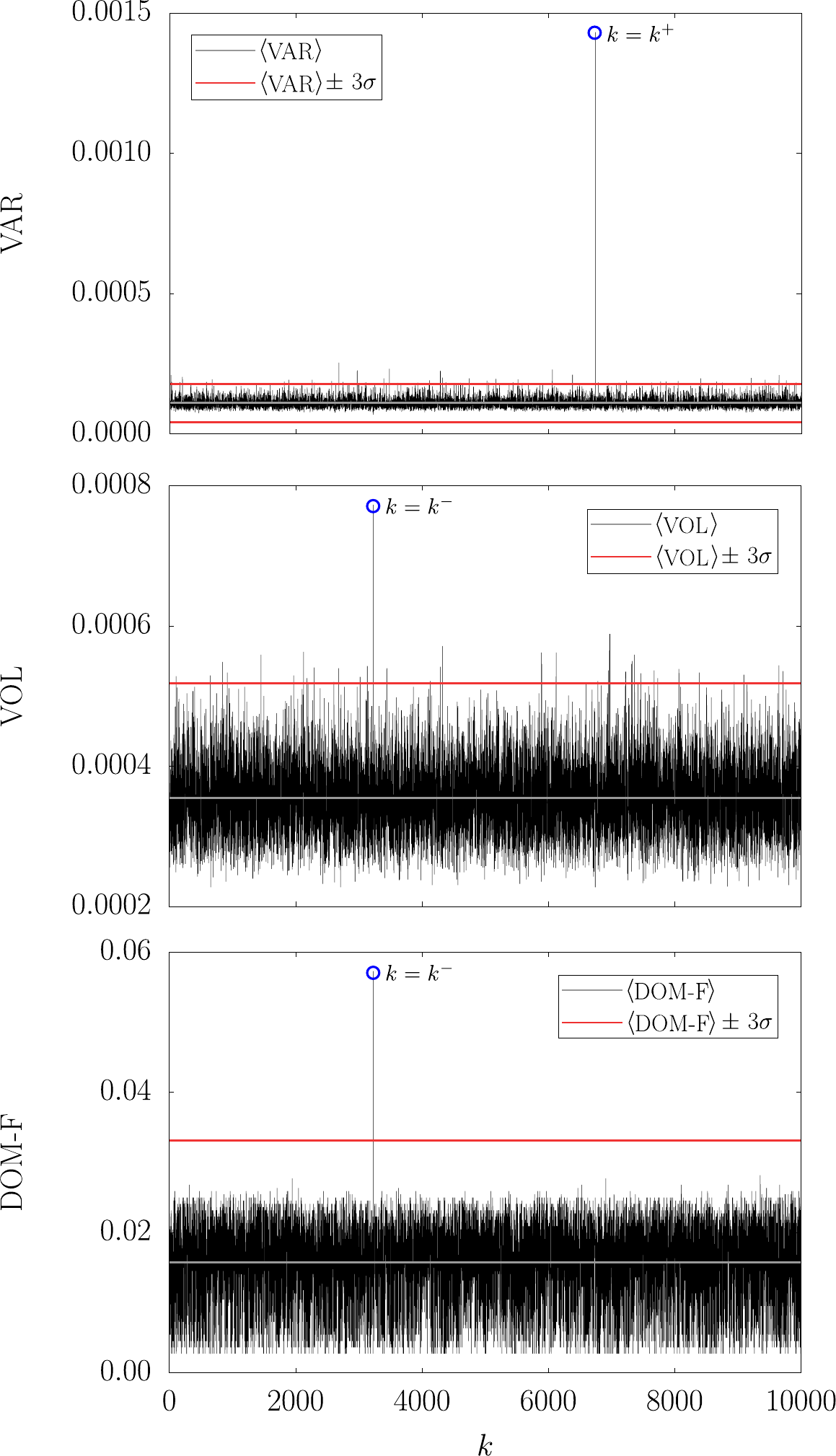}
    \caption{Distribution of three different time-series measures, Variance (VAR), Volatility (VOL), and Dominant Frequency (DOM-F), computed across the same ensemble of $10^4$ distinct simulations of the RPS model as in Fig.~\ref{fig:dist_1} and \ref{fig:dist_10}. Outlier simulations, detected via \( \langle Q \rangle \pm 3\sigma \), are highlighted with blue circles.}
    \label{fig:dist_ts}
\end{figure}

In this appendix, we present a complementary analysis aimed at comparing the recurrence-based quantifiers explored throughout the main text with conventional (linear) scalar measures derived directly from the time-series, specifically in the task of detecting outlier simulations presented in Sec.\ \ref{subsec:mobility_detect}. While recurrence quantifiers capture the geometric and temporal structures within the analysed dynamics, traditional time-series analysis focuses on basic signal properties that might, in some cases, perform as well as their recurrence-based counterparts. To explore this comparison, we computed three standard scalar indicators for each simulation in the ensemble, namely the variance (VAR), volatility (VOL), and the dominant frequency (DOM-F) obtained via Fourier analysis, following the same construction and presentation adopted in Sec.~\ref{subsec:mobility_detect}.

Figure~\ref{fig:dist_ts} presents the ensemble distributions of the three selected time-series quantifiers. In general terms, each measure captures a distinct aspect of the signal’s properties: VAR reflects the overall dispersion of values around the mean; VOL estimates the average rate of change through the mean absolute derivative; and DOM-F identifies the dominant frequency component via the Fourier spectrum. Across the ensemble, these quantifiers exhibit relatively narrow distributions, with only a few simulations identified as outliers beyond the $3\sigma$ threshold. Notably, each scalar tends to highlight different simulations as outliers, indicating that they are indeed sensitive to distinct signal features. However, the limited variability observed, particularly for VAR and VOL, suggests that these measures may offer only coarse discrimination across the ensemble, especially when compared to the recurrence-based results discussed in Sec.~\ref{subsec:mobility_detect}.

Indeed, while compared to the recurrence-based methodology, the linear time-series quantifiers were able to detect only one of the two main outliers in the ensemble. In contrast, the recurrence-based analysis successfully identified both $k^-$ and $k^+$ as outliers, thus providing a more complete picture of the ensemble's deviations. This was particularly evident in the quantifiers DET, LAM, ENTR-L, and ENTR-V (Figs.~\ref{fig:dist_1} and~\ref{fig:dist_10}), all of which marked these simulations beyond the $3\sigma$ threshold. This reinforces the idea that recurrence-based measures are more effective in capturing subtle and structurally relevant differences in the dynamics, features that may be overlooked by traditional, linear time-series descriptors.

\bibliographystyle{apsrev4-2}
%


\end{document}